\documentclass[preprint,showpacs]{revtex4}
\usepackage{amsfonts}
\usepackage{amsmath}
\usepackage{amssymb}
\usepackage{graphicx}
\usepackage{hyperref}
\usepackage{color}
\usepackage{epsfig}

\begin{document}

\title{Can the Reissner-Nordstr\"{o}m black hole or Schwarzschild black hole
be the stable Planck-scale particle accelerator?}
\author{Yi Zhu$^{1}$, Shao-Feng Wu$^{1,2}$\thanks{%
Corresponding author. Email: sfwu@shu.edu.cn; Phone: +86-021-66136202.}, Yin
Jiang$^{1,2}$, and Guo-Hong Yang$^{1,2}$}

\begin{abstract}
It is shown that the extremal Reissner-Nordstr\"{o}m black hole, the
non-extremal one with multiple scattering particles, and the Schwarzschild
black hole with radial head-on particles are stable under the collision of
the particles near the horizon, if the back-reaction effect and the effect
generated by gravity of particles are involved. Moreover, the collision near
Reissner-Nordstr\"{o}m black holes with astrophysically typical mass can not
generate the Planck-scale center-of-mass energy. However, the head-on
collision near the typical primordial black hole could just occur at the
Planck-energy scale.
\end{abstract}

\affiliation{$^{1}$Department of physics, Shanghai University, Shanghai, 200444, P. R.
China}
\affiliation{$^{2}$The Shanghai Key Lab of Astrophysics, Shanghai, 200234, P. R. China}
\maketitle

\section{Introduction}

Recently, Ba\~{n}ados, Silk and West (BSW) \cite{BSW Mechanic} proposed a
mechanism to obtain arbitrary high center-of-mass (CM) energy of two
particles collided on the horizon of an extremal Kerr black hole (BH), which
was hence asserted as a natural Planck-scale particle accelerator and the
possible origin for the very highly energetic astrophysical phenomena.
However, the authors of \cite{Comment} and \cite{Jacobson} pointed out that
some factors, such as the astrophysical limit of maximal spin of Kerr BHs,
would inhibit the Plank-scale collision.

To avoid the astrophysical limitation of Kerr BHs, several methods have been
presented. The first is the called multiple scattering mechanism, by which
the BSW process still can be used to get arbitrary high CM energy of
colliding particles near the nonextremal Kerr BH \cite{pavlov}. Another more
direct idea is to consider different extreme rotating BHs \cite{KN}.
Actually, Zaslavskii showed the unbound energy of colliding particles on the
general rotating extreme horizon or the nonextremal horizon considering
multiple scattering \cite{general rotating}. Noticing the alternative
options for generating extremal black holes, the BSW mechanism has been
further studied by calculating the escaping flux of massless particles for
maximally rotating black holes, and it was suggested that the received
spectrum should typically contain signatures of highly energetic products
\cite{Banados}, see also the sequent numerical estimation \cite{Williams}.

Most works \cite{Sen,Jie Yang,ISCO,naked
singularity,Chen,cylindrical,inner,Said} on the BSW mechanism necessitate
the rotation of BHs. But in \cite{RN}, it was proposed that a non-rotating
but charged Reissner-Nordstr\"{o}m (RN) BH can also serve as an accelerator
with arbitrarily high CM energy of charged particles collided at the extreme
horizon or nonextremal horizon considering the multiple scattering. In
particular, it was demonstrated that the upper bound of the electric charge
of BHs after Schwinger emission is large enough to allow the ultra-high CM
energy of charged colliding particles, provided that the nonextremal BH is
not too light ($>10^{20}g$). Furthermore, the general stationary charged BH
was studied in detail \cite{our paper}, suggesting that the potential
acceleration to large energies should be taken seriously as the
manifestation of general properties of rotating or charged BHs. Other
general kinematic explanation was presented in \cite{kinematic explanation}.

All the aforementioned works only considered the collision between ingoing
particles. The collision between ingoing and outgoing particles has been
studied in \cite{Piran} very early, and recently reintroduced more generally
in \cite{general explanation}, where it was shown that the unlimited CM
energy can also be attained. Compared with the BSW process that requires one
particle having critical angular momentum or charge, this divergence can be
seen as a direct consequence of infinite redshift near the horizon.

However, beside the astrophysical limit of Kerr BHs, there are other factors
to prohibit the ultra-energetic collisions near extreme Kerr BHs, such as
the gravitational radiation \cite{Comment}, the back-reaction effect \cite%
{Comment}, and the effect of gravity generated by colliding particles \cite%
{shell}. Here the called back-reaction effect is considered as due to
absorbing the first pair of colliding particles, then the extremal Kerr BH
becomes non-extremal and the CM energy declines rapidly. The effect of
gravity of particles seems similar to the back-reaction effect in its
physical significance, but is tackled more rigorously in the case of the
collision of two spherical dust shells in the neighborhood of an extreme RN
BH. It was shown \cite{shell} that an upper limit exists for the total
energy of colliding shells in the CM frame in the observable domain due the
gravity of shells. Since the similarity between the BSW process for RN\ BHs
and Kerr BHs, it was argued that the upper limit also exists for extreme
Kerr BHs.

In this paper, we will take into account the back-reaction effect and the
effect arising from the gravity of particles (shells) for the collision of
particles near the extremal RN BH, the collision near the non-extremal RN BH
considering multiple scattering, and the radial head-on collision near the
Schwarzschild BH. The gravitational radiation can be neglected since these
cases are spherically symmetric. We will focus on two important problems:
whether or not the CM energy of colliding particles can reach the Planck
scale and even the BH mass. Note that the later problem means whether the BH
is stable against minor perturbations induced by dropping small particles
into the black hole. This new paradoxical instable mechanism of BHs under
the BSW process was suggested by Lake \cite{Lake} and clearly presented in
\cite{shell}, where the extremal RN BH was found to be stable if the gravity
of particles is incorporated. We will study this instability for other
mentioned cases.

\section{Back-reaction effects}

We set out to investigate back-reaction effects. The metric for RN
spacetimes is%
\begin{equation}
ds^{2}=-fdt^{2}+f^{-1}dr^{2}+r^{2}(d\theta ^{2}+\sin ^{2}\theta d\phi ^{2}),
\end{equation}%
where%
\begin{equation*}
f=1-\frac{2M}{r}+\frac{Q^{2}}{r^{2}}.
\end{equation*}%
The event horizon is located at $r_{H}=M+\sqrt{M^{2}-Q^{2}}$. Considering a
test particle with charge $q$ and energy $E$ per unit rest mass $m$, the
radial timelike geodesics is
\begin{equation}
u^{t}\equiv \frac{dt}{d\tau }=\frac{1}{f}\left( E-\frac{qQ}{r}\right)
,\;u^{r}\equiv \frac{dr}{d\tau }=\pm \sqrt{\left( E-\frac{qQ}{r}\right)
^{2}-f},  \label{ut}
\end{equation}%
where the positive and minus signs of $u^{r}$ correspond to outgoing and
ingoing geodesics, respectively. The CM energy can be obtained by%
\begin{equation}
E_{c.m.}^{2}=-(m_{1}u_{1}+m_{2}u_{2})^{2},  \label{EC0}
\end{equation}%
where index $i=1,2$ denote two different particles.

\subsection{Extremal RN BHs}

For two test particles with the equal mass $m_{0}$, charges $q_{i}\;$($i=1,2$%
), and energy parameters $E_{i}$ per unit rest mass, the CM energy (\ref{EC0}%
) can be expressed as \cite{RN}
\begin{equation}
\frac{E_{c.m.}^{2}}{2m_{0}^{2}}=1+\frac{1}{2}\left( \frac{E_{2}r_{H}-Qq_{2}}{%
E_{1}r_{H}-Qq_{1}}+\frac{E_{1}r_{H}-Qq_{1}}{E_{2}r_{H}-Qq_{2}}\right)
\label{CM1}
\end{equation}%
Consider extremal RN BHs with $Q=M$ and one ingoing particle with critical
charge $q_{1}=E_{1}$, which is capable of touching the extremal horizon from
infinity. It is easy to notice that the CM energy could be divergent when $%
q_{1}=E_{1}$ and $q_{2}\neq E_{2}$, which implies that the collision on the
extremal horizon of RN BHs might produce arbitrary high CM energy, and seems
to destroy the BH background. We note that the angular momentum of particles
will not lead to the qualitative difference.

\begin{figure}[tbp]
\includegraphics[width=7cm]{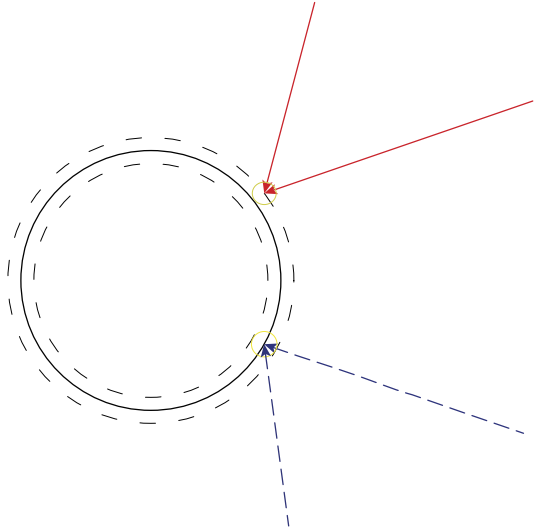}
\caption{Schematic diagram of the back-reaction effect of extremal RN BHs.
The blue dashed lines represent the first pair of colliding particles. They
collide on the horizon of extremal RN BHs, which is denoted by the black
solid circle. After absorbing the first pair of colliding particles, the
degenerated horizon is split into the event horizon and Cauchy horizon which
are represented by the black dashed circles. The second pair of colliding
particles are represented by the red solid lines, they collide on the event
horizon.}
\label{RN BH}
\end{figure}

Now we will tackle the back-reaction effect following the spirit in Ref.
\cite{Comment}, but not using the dimensionless parameters of BHs. Upon
absorption of the first pair of colliding particles, the new charge, mass
and horizon of BHs are (see Fig. \ref{RN BH})%
\begin{equation}
Q^{\prime }=Q+m_{0}(q_{1}+q_{2}),\;M^{\prime
}=M+m_{0}(E_{1}+E_{2}),\;r_{H}^{\prime }=M^{\prime }+\sqrt{M^{\prime
}-Q^{\prime 2}}\text{.}  \label{QM}
\end{equation}%
Since $\sigma =\frac{m_{0}}{M}\ll 1$, the CM energy (\ref{CM1}) after the
first collusion can be obtained as%
\begin{equation}
\frac{E_{c.m.}^{2}}{2m_{0}^{2}}=1+\frac{E_{2}-q_{2}}{2E_{1}\sqrt{%
2E_{1}+4E_{2}-2q_{2}}}\frac{1}{\sqrt{\sigma }}+\mathcal{O}\left( \sigma
\right) ^{0}.  \label{CM2}
\end{equation}%
Focusing on the order of magnitude of the CM\ energy, Eq. (\ref{CM2}) can be
recast as%
\begin{equation}
E_{c.m.}\lesssim m_{0}^{\frac{3}{4}}M^{\frac{1}{4}}\simeq 10^{12}\left(
\frac{m_{0}}{1MeV}\right) ^{\frac{3}{4}}\left( \frac{M}{100M_{\odot }}%
\right) ^{\frac{1}{4}}GeV.  \label{ERN}
\end{equation}%
We note that Eq. (\ref{ERN}) is same as the order of magnitude of CM energy
obtained in \cite{Comment} for extremal Kerr BHs. Thus, one can have the
similar result, namely, the extremal BHs are stable under the particle
collision as long as the mass of the particles is smaller than the mass of
the BHs. In addition, the Planck energy is hard to be attained for typical
values of parameters. For instance, for the BH with $M\sim 100M_{\odot }$, a
hypothetical dark matter particle would need a mass $m_{0}\sim 10^{10}MeV$
to allow for more than one Planck-scale event. The collision of a single
electron pair would reduce the charge of a $100M_{\odot }$ BH sufficiently
to inhibit any further Planck-scale collision.

\subsection{Non-extremal RN BHs}

We have considered the extremal\emph{\ }RN BHs with the mass of typical
astrophysical BHs $M\sim 100M_{\odot }$, which is only of academic interest
since the astrophysical BHs can not possess a large charge and should be
non-extremal. However, in the non-extremal RN background, a particle with
the critical charge $q_{1}=E_{1}r_{H}/Q$ can not arrive at the event horizon
from infinity \cite{RN,our paper}. Similar situation happens in non-extremal
Kerr BHs where the particle with critical angular momentum can not reach the
horizon. However, noticing that there exists some region close to the
horizon where one has particles moving with angular momentum arbitrary close
to the critical value, Pavlov \cite{pavlov} presented that the BSW process
is also applicable to non-extremal Kerr BHs by invoking the multiple
scattering mechanism to amplify the angular momentum of the falling particle
from infinity. This mechanism has been used to RN BHs \cite{RN} and more
general charged background \cite{our paper} where the small charge of
particle from infinity can be amplified near the horizon. The multiple
scattering mechanism in charged BHs is effective since a particle with the
charge close to the critical value $q_{1}=(1-\delta )E_{1}r_{H}/Q$, which is
generated by the multiple scattering near the horizon, may exist near the
event horizon \cite{RN,our paper}. The CM energy of the collision between a
near critical particle and another particle with charge $q_{2}$ on the event
horizon is
\begin{equation}
\frac{E_{c.m.}^{2}}{2m_{0}^{2}}=\frac{E_{2}r_{H}-Qq_{2}}{2E_{1}r_{H}}\frac{1%
}{\delta }+\mathcal{O}\left( \delta \right) .  \label{MS energy}
\end{equation}%
One can find that the CM energy will be divergent while $\delta \rightarrow
0 $.

Now, let us count the back-reaction effect. Due to absorbing the first pair
of colliding particles, the parameters of BHs become to Eq. (\ref{QM}),
which are similar to the case of extremal RN BHs. However, because the
multiple scattering mechanism is still applicable for the following
collisions to provide the particle with new near critical charge $%
q_{1}^{\prime }=(1-\delta )E_{1}r_{H}^{\prime }/Q^{\prime }$, the
corresponding CM energy will be the same as Eq. (\ref{MS energy}), except to
replace $r_{H}$ by $r_{H}^{\prime }$. Thus, the CM energy of multiple
scattering particles colliding near the non-extremal RN BH will\ still be
divergent after involving the back-reaction effect. We also note that, by a
similar consideration for the non-extremal Kerr BHs with multiple scattering
particles, one can obtain the same conclusion. Thus, if one would like to
guarantee the stability of BHs under the BSW process considering multiple
scattering, as it should be, other effects need to be considered.

\subsection{Radial head-on collision in Schwarzschild spacetimes}

The authors of Ref. \cite{BSW Mechanic} showed that one could not get
ultra-high CM energy by the collision of two ingoing particles in
Schwarzschild spacetimes. However, Zaslavskii noticed that \cite{general
explanation} the unbound CM energy can be achieved by the collision at
horizon between outgoing particle and ingoing particle. In this subsection,
we will calculate the back-reaction effect for the radial head-on collision.
Here, the back-reaction effect has somewhat of an inverse relationship to
the one discussed above, where it is generated by the BH absorbing the
colliding particles. Now the back-reaction effect is considered as arising
from the shrink of the horizon due to emitting an outgoing particle by BHs.
\begin{figure}[tbp]
\includegraphics[width=7cm]{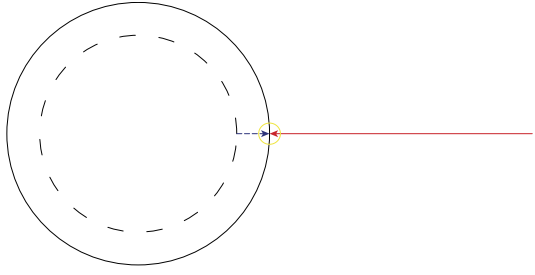}
\caption{Schematic diagram of the back-reaction effect for the radially
head-on collision in Schwarzschild spacetimes. The solid circle represents
the location of horizon before the collision. It shrinks back to the dashed
circle after emitting a particle, which is represented by the blue dashed
line. The outgoing particle collides with an ingoing particle, which is
represented by the red solid line, on the location of the original horizon.}
\label{outgoing-ingoing}
\end{figure}

According to Eqs. (\ref{ut}) and (\ref{EC0}) with $Q=q_{i}=0$, one can
obtain the CM energy of the collision between an outgoing particle and an
ingoing particle in Schwarzschild BHs, which is%
\begin{equation}
E_{c.m.}^{2}=m_{1}^{2}+m_{2}^{2}+2m_{1}m_{2}\frac{E_{1}E_{2}r+\sqrt{\left(
E_{1}^{2}r-r+2M\right) \left( E_{2}^{2}r-r+2M\right) }}{r-2M}.  \label{CM3}
\end{equation}%
When the collision occurs on the horizon, i.e. $r\rightarrow 2M$, the
denominator of Eq. (\ref{CM3}) equates to zero and the numerator is
nonvanishing. Hence, this CM energy of the collision is divergent on the
horizon.

To take the back-reaction effect into account, we will assume that the
original mass of BH is $M+E_{2}m_{2}$ and the horizon is $r_{H}=2\left(
M+E_{2}m_{2}\right) $. After the BH emits a particle with mass $m_{2}$ and
energy $E_{2}$, the horizon will be changed to $r_{H}=2M$ and the collision
should occur at $r\geqslant 2(M+E_{2}m_{2})$, see Fig. \ref{outgoing-ingoing}%
. So the CM energy (\ref{CM3}) is less than that of the collision at $%
r=2(M+E_{2}m_{2})$:
\begin{equation}
E_{c.m.}^{2}=\frac{4m_{1}m_{2}E_{1}E_{2}}{\sigma }+\mathcal{O}\left( \sigma
\right) ^{0}=4E_{1}E_{2}m_{1}M+\mathcal{O}\left( \sigma \right) ^{0}
\label{CM7}
\end{equation}%
where $\sigma =m_{2}/M$. From this equation, we can estimate the order of
magnitude for the CM energy as following:
\begin{equation}
E_{c.m.}\lesssim m_{1}^{\frac{1}{2}}M^{\frac{1}{2}}\simeq 10^{28}\left(
\frac{m_{1}}{1MeV}\right) ^{\frac{1}{2}}\left( \frac{M}{100M_{\odot }}%
\right) ^{\frac{1}{2}}GeV.  \label{CM8}
\end{equation}%
Several remarks are in order. First, the colliding CM energy does not relate
to the rest mass of outgoing particles. This suggests that Eq. (\ref{CM8})
is general even for the collision between an outgoing photon and an ingoing
massive particle. Actually, one can check it following the similar process,
although the geodesics of photons are different to massive particles.
Second, similar to the previous case of extremal RN BHs, the colliding CM
energy can not reach the order of magnitude of BH's mass, since it needs the
rest mass of ingoing particle $m_{1}\sim M$. Thus, the Schwarzschild BH is
stable under the collision of outgoing and ingoing particles. Third, the CM
energy is proportional to $M^{1/2}$, different from $M^{1/4}$ in Eq. (\ref%
{ERN}). This is an essential difference. Consequently, for a Schwarzschild
BH with the $100$ solar mass and $m_{1}\sim 1MeV$, the CM energy is $%
E_{c.m.}\sim 10^{28}GeV$, which is higher than the Planck energy. Finally,
let us discuss how to set an out-going particle near the horizon. One might
consider a particle which starts at infinity and falls into the horizon. In
principle, it is possible to arrange that the particle decays into two
particles near the horizon, where one particle falls into the horizon and
the other goes out. The process is similar to the Penrose process with an
additional restriction that the split just occurs outside the event horizon.
So it would be quite rare in astrophysics. To be more realistic, we would
like to invoke the mechanism of Hawking radiation. Consider vacuum
fluctuations which create a particle-antiparticle pair in the vicinity of
the horizon of a black hole. If the region is sufficiently small (that is
just necessary for the collision with high CM energy), an antiparticle with
negative energy can be created inside the horizon and an escaping particle
outside with finite positive energy. However, the Hawking temperature for
Schwarzschild BHs is $T\sim 10^{-7}\frac{M_{\odot }}{M}$K which is very low
for astrophysically large BHs and they would evaporate completely on a very
long time scale $\tau \sim 10^{64}\left( \frac{M}{M_{\odot }}\right) ^{3}$%
yr. Thus, we are turned to the small primordial black holes with the typical
mass $M\sim 10^{15}g\sim 10^{-18}M_{\odot }$, which are hot enough ($\sim
10^{11}$K) to radiate a significant fraction of their mass over the
14-billion-year age of the universe and would be exploding today \cite{Carr}%
. Interestingly, one can find that the small mass of such primordial black
holes can just afford the Planck-scale collision.

\section{The gravity generated by colliding shells}

We will investigate the effect of the gravity generated by colliding
particles. The approximate method of evaluating this effect is to consider
the collision of two infinitesimal thin spherical shells. In Ref. \cite%
{shell}, it has been showed that the CM energy of this collision in the
extremal RN background can not reach the Plank energy respecting this
effect. Here, we would like to study the colliding energy for
infinitesimally thin spherical shells in nonextremal RN spacetimes and the
radial head-on collision in Schwarzschild spacetimes.

\begin{figure}[tbp]
\includegraphics[width=7cm]{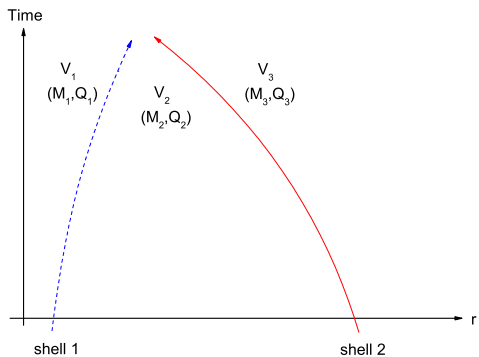}
\caption{Schematic spacetime diagram of two spherical timelike shells. Two
shells divides the spacetime into three parts. Shell 1 (blue dash line) is
outgoing and shell 2 (red solid line) is ingoing. For the case with two
ingoing shells, see Fig. 2 in Ref. \protect\cite{shell}.}
\label{shells}
\end{figure}

The existence of spherical shells divides the spacetime into three regions,
as show in Fig. \ref{shells}, where we denote them with the index $i=1,2,3$.
According to the Birkhoff's theorem, the metrics of those spacetimes can be
described by
\begin{equation}
ds^{2}=-f_{i}dt_{i}^{2}+f_{i}^{-1}dr^{2}+r^{2}(d\theta ^{2}+\sin ^{2}\theta
d\phi ^{2}),
\end{equation}%
where $f_{i}=1-\frac{2M_{i}}{r}+\frac{Q_{i}{}^{2}}{r^{2}}$, no matter the
spherical objects are radially collapsing or exploding. Note that all
coordinates except for the time $t$ are common to different regions. The BH
horizons in different regions are located at $r_{i}=M_{i}+\sqrt{%
M_{i}{}^{2}-Q_{i}^{2}}$.

\subsection{RN spacetimes}

Consider that the shell 1 (inner shell) is a charged dust shell with charge $%
q_{1}=Q_{2}-Q_{1}$ and $\epsilon \mu =M_{2}-M_{1}$, where $\mu $ is the
proper mass of the shell and $\epsilon $ is the specific energy; the shell 2
(outer shell) is composed of neutral dust and has the same proper mass and
specific energy as the shell 1. Using Israel junction conditions \cite%
{Israel}, the effective potential of shell 1 has been obtained in \cite%
{shell}:%
\begin{equation*}
V_{eff}^{1}=-\left( u_{1}^{r}\right) ^{2},
\end{equation*}%
where the radial component of 4-velocity is written in the form%
\begin{equation}
u_{1}^{r}=-\sqrt{-1-\frac{q_{1}^{2}}{4r^{2}}-\frac{\langle Q\rangle _{1}^{2}%
}{r^{2}}+\frac{2\langle M\rangle _{1}}{r}+\left( \epsilon -\frac{%
q_{1}\langle Q\rangle _{1}}{r\mu }\right) ^{2}+\frac{\mu ^{2}}{4r^{2}}},
\label{shell velocity}
\end{equation}%
with
\begin{equation}
\langle M\rangle _{1}=\frac{M_{2}+M_{1}}{2},\quad \langle Q\rangle _{1}=%
\frac{Q_{2}+Q_{1}}{2}.  \label{V1}
\end{equation}%
For shell 2, the radial velocity has the same form except to replace $q_{1}$
as zero and take $\langle M\rangle _{1}$ and $\langle Q\rangle _{1}$ as
\begin{equation}
\langle M\rangle _{2}=\frac{M_{3}+M_{2}}{2},\quad \langle Q\rangle _{2}=%
\frac{Q_{3}+Q_{2}}{2}.  \label{V2}
\end{equation}%
Introducing the \textquotedblleft center of mass frame\textquotedblright\ at
the collision event of the shells, the CM energy of the colliding shells has
been given by \cite{shell}%
\begin{equation}
\frac{E_{c.m.}^{2}}{2\mu ^{2}}=1-g_{ab}u_{1}^{a}u_{2}^{b},  \label{CM4}
\end{equation}%
which can be rewritten as%
\begin{equation}
\frac{E_{c.m.}^{2}}{2\mu ^{2}}=1-\frac{u_{1}^{r}u_{2}^{r}}{f_{2}}+\sqrt{%
\left[ 1+\frac{\left( u_{1}^{r}\right) ^{2}}{f_{2}}\right] \left[ 1+\frac{%
\left( u_{2}^{r}\right) ^{2}}{f_{2}}\right] }.  \label{CM5}
\end{equation}%
Eq. (\ref{CM5}) can be expanded as%
\begin{equation}
\frac{E_{c.m.}^{2}}{2\mu ^{2}}=\frac{\left( u_{1}^{r}+u_{2}^{r}\right) ^{2}}{%
2u_{1}^{r}u_{2}^{r}}+\mathcal{O}\left( f_{2}\right) ^{1}.  \label{CM9}
\end{equation}%
At first glance, the CM energy could be divergent when $f_{2}=0$, $\left(
u_{1}^{r}\right) ^{2}=0$ and $\left( u_{2}^{r}\right) ^{2}\neq 0$. However,
noticing that $r_{3}$, i.e. the event horizon in region 3, is larger than
the $r_{2}$, distant observers could not see the collision on $r_{2}$.
Consequently, the observable CM energy is less than that of the collision at
$r_{3}$.

\begin{figure}[tbp]
\includegraphics[width=7cm]{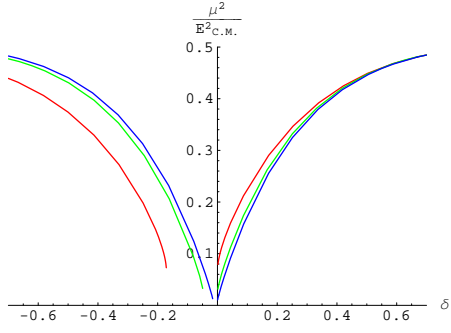}
\caption{The inverse of the observable CM energy $2\protect\mu ^{2}$/$%
E_{c.m.}^{2}$ with respect to the charge of inner shell $q_{1}=q_{c}(1-%
\protect\delta )$. The charge of BHs is assumed as $x=0.5$ and specific
energy $\protect\epsilon =2\,$. The red, green, and blue lines (from top to
bottom in the region $\protect\delta >0$) denote $\protect\sigma =10^{-2}$, $%
10^{-3}$, $10^{-4}$, respectively. This figure indicates $q_{c}$ approaching
the critical charge for the observable maximized CM energy when $\protect%
\sigma \rightarrow 0$.}
\label{qc}
\end{figure}

Now let us calculate the observable maximized CM energy. Since $\sigma =\mu
/M_{1}$ is very small, $f_{2}(r_{3})$ is close to $f_{2}(r_{2})=0$. Then Eq.
(\ref{CM9}) at $r_{3}$ is still effective in general, except for the case of%
\begin{equation}
\left[ u_{1}^{r}(r_{3})\right] ^{2}=0,\;\left[ u_{2}^{r}(r_{3})\right]
^{2}\neq 0.  \label{con}
\end{equation}%
Eq. (\ref{con}) means%
\begin{equation}
q_{1}=q_{c}\equiv Q_{2}-\sqrt{Q_{2}^{2}-2r_{3}\epsilon \mu +\mu ^{2}-2\mu
\sqrt{Q_{2}^{2}+r_{3}(r_{3}-2M_{1}-2\epsilon \mu )}}.  \label{q1}
\end{equation}%
Note $\left[ u_{2}^{r}(r_{3})\right] ^{2}\neq 0$ since%
\begin{equation*}
\left[ u_{2}^{r}(r_{3},q_{c})\right] ^{2}=\epsilon ^{2}+\mathcal{O}\left(
\sigma \right) ^{1}.
\end{equation*}%
An important observation is that the divergent Eq. (\ref{CM9}) suggests $%
q_{c}$ approaching the critical charge for the observable maximized CM
energy when $f_{2}(r_{3})$ approaches zero. In other word, we can treat $%
q_{c}$ as the critical charge since $\sigma $ is very small. To confirm that
the derivation of critical charge is rigorous, we have checked it using
numerical method, see Fig. \ref{qc} for Eq. (\ref{CM5}) at $r_{3}$ with
respect to $q_{1}$. Note that we have set $Q_{2}=M_{2}(1-x)$ ($0\leq x\leq 1$%
) and $q_{1}=q_{c}(1-\delta )$ ($\delta \leq 1$) to characterize the
deviation.

Furthermore, taking $r=r_{3}(1+y)$ ($y\geq 0$), one can find that the shell
1 with $q_{c}$ can not touch the $r_{3}$ from infinity since the effective
potential with $q_{1}=q_{c}$ is%
\begin{equation*}
V_{eff}^{1}=\frac{2\sqrt{x(2-x)}\left[ 1-\sqrt{x(2-x)}\right] }{(1-x)^{2}}y+%
\mathcal{O}\left( y\right) ^{2}+\mathcal{O}\left( \sigma \right) ^{\frac{1}{2%
}}>0.
\end{equation*}%
However, following the spirit in \cite{pavlov}, one might consider that the
shell 1 is generated by the multiple scattering near $r_{3}$ with near
critical charge $q_{1}=q_{c}(1-\delta )$, since the effective potential is
negative at $r_{3}$:%
\begin{equation}
V_{eff}^{1}=-4\delta ^{2}+\mathcal{O}\left( \delta \right) ^{3}+\mathcal{O}%
\left( \sigma \right) ^{\frac{1}{2}}<0.  \label{Veff}
\end{equation}%
Moreover, one should be noticed that in order that the outer shell overtakes
the inner shell, $\left[ u_{2}^{r}(r_{3})\right] ^{2}-\left[ u_{1}^{r}(r_{3})%
\right] ^{2}$ should be positive, which can be seen from%
\begin{equation*}
\left[ u_{2}^{r}(r_{3})\right] ^{2}-\left[ u_{1}^{r}(r_{3})\right]
^{2}=\epsilon ^{2}+\mathcal{O}\left( \sigma \right) ^{\frac{1}{2}}+\mathcal{O%
}\left( \delta \right) ^{1}.
\end{equation*}

Thus, the maximized CM energy can be evaluated from Eq. (\ref{CM5}) with
critical charge $q_{1}=q_{c}(1-\delta )$ at $r_{3}$:%
\begin{equation}
\frac{E_{c.m.}^{2}}{2\mu ^{2}}=\sqrt{\frac{\epsilon }{2}}\sqrt{1+\sqrt{x(2-x)%
}}\frac{1}{\sqrt{\sigma }}+\mathcal{O}\left( \sigma \right) ^{0}+\mathcal{O}%
\left( \delta \right) ^{\frac{1}{2}}.  \label{CM6}
\end{equation}%
Since Eq. (\ref{CM6}) has the same order of magnitude of Eq. (\ref{CM2}), we
know that the CM energy declines rapidly upon counting the gravity of
shells, and it can not reach the Planck scale.

We note that when $x=0$ denoting region 2 as an extremal RN spacetime, Eq. (%
\ref{CM6}) reduces to%
\begin{equation*}
\frac{E_{c.m.}^{2}}{2\mu ^{2}}=\sqrt{\frac{\epsilon }{2}}\frac{1}{\sqrt{%
\sigma }}+\mathcal{O}\left( \sigma \right) ^{0},
\end{equation*}%
which is a little larger than the result obtained in \cite{shell}%
\begin{equation*}
\frac{E_{c.m.}^{2}}{2\mu ^{2}}=\sqrt{\frac{\epsilon }{2}}(\epsilon -\sqrt{%
\epsilon ^{2}-1})\frac{1}{\sqrt{\sigma }}+\mathcal{O}\left( \sigma \right)
^{0},
\end{equation*}%
where the involved critical charge $q_{c}$ is not exactly solved from $\left[
u_{1}^{r}(r_{3})\right] ^{2}=0$ but from $\left[ u_{1}^{r}(r_{2})\right]
^{2}=0$.

\subsection{The head-on collision}

For the radial head-on collision in Schwarzschild spacetimes, equations of
motion for the shells are described by Eqs. (\ref{shell velocity}), (\ref{V1}%
) and (\ref{V2}) with $q_{i}=Q_{i}=0$, but the radial component of
4-velocity of inner shell should add a minus sign. Thus, one can calculate
the CM energy of the head-on collision of shells from Eq. (\ref{CM5}), which
is given by%
\begin{eqnarray*}
\frac{E_{c.m.}^{2}}{2\mu ^{2}} &=&\frac{-1}{4r(r-r_{2})}\bigg[%
8M_{1}r-4r^{2}\left( \epsilon ^{2}+1\right) +8r\epsilon \mu +\mu ^{2}- \\
&&\sqrt{8M_{1}r+4r^{2}\left( \epsilon ^{2}-1\right) +12r\epsilon \mu +\mu
^{2}}\sqrt{8M_{1}r+\left[ 2r\left( \epsilon -1\right) +\mu \right] \left[
2r\left( \epsilon +1\right) +\mu \right] }\bigg].
\end{eqnarray*}%
It is divergent when $r=r_{2}$. However, the observable CM energy should be
less than that of the collision at $r_{3}$, which is%
\begin{equation*}
\frac{E_{c.m.}^{2}}{2\mu ^{2}}=\frac{2\epsilon }{\sigma }+\mathcal{O}\left(
\sigma \right) ^{0}.
\end{equation*}%
This equation is similar to Eq. (\ref{CM7}), indicating that a Schwarzschild
BH could be a stable Planck-scale particle accelerator when gravity of
shells is involved.

\section{Conclusion}

In this work, we show that the BSW process involving the back-reaction
effect can not produce the Planck-scale CM energy in the background of the
extremal RN BH with astrophysically typical mass, but for the non-extremal
RN BH, the CM energy of colliding particles generated by multiple scattering
still can be arbitrary high. However, if the effect of the gravity of
particles (shells) is incorporated, even the Planck energy can not be
attained. This case demonstrates that the back-reaction effect due to
absorbing the first pair of colliding particles and the effect generated by
gravity of particles sometimes are different operationally, although their
physical significance seems similar. Moreover, it is found that the
Schwarzschild BH is stable under the radial head-on collisions if any effect
is involved. Interestingly, for small primordial BHs with the typical mass $%
10^{15}g$, the head-on collision could just occur at the Planck-energy
scale. Whether this result might have the astrophysically observable
phenomenon is very deserved to be investigated in the future.

\begin{acknowledgments}
SFW was partially supported by National Natural Science Foundation of China
(No. 10905037). This work was partially supported by Shanghai Leading
Academic Discipline Project No. S30105 and the Specialized Research Fund for
the Doctoral Program of Higher Education under No. 20093108110004.
\end{acknowledgments}

\end{document}